\newcommand{\arcsec}{$^{\prime\prime}$}
\newcommand{\arcmin}{$^{\prime}$}

\newcommand{\kms}{${\rm\ km\ s}^{-1}$}

\newcommand{\pccm}{{\rm\ cm}$^{-3}$}

\newcommand{\etal}{{et al.\,}}

\newcommand{\ha}{H$\alpha$}
\newcommand{\us}{_{\rm s}}

\newcommand{\uj}{_{\rm j}}

\def\u#1{_{\rm #1}}
\def\r#1{{\rm #1}}
\newcommand{\mpr}{\r{m_p}}

\def\ee#1{\times10^{#1}}
\newcommand{\msol}{M_\odot}

\documentstyle[12pt,aaspp4,flushrt]{article}

\lefthead{Jets in the giant envelope of KjPn 8}

\righthead{ Steffen \& L\'opez}

\begin{document}

\title{Jets and the shaping of the giant bipolar envelope of the 
planetary nebula KjPn 8}

\author{ W. Steffen}
\affil{Department of Physics and Astronomy, University of Manchester,
Oxford Rd., Manchester, M13 9PL, UK}
\author{J. A. L\'{o}pez,}
\affil{Instituto de Astronom\'{\i}a, UNAM, Apdo. Postal 877,
Ensenada, B.C. 22800, M\'{e}xico.}

\begin{abstract}
 
A hydrodynamic model involving cooling gas in the stagnation region of
a collimated outflow is proposed for the formation of the giant
parsec-scale bipolar envelope that surrounds the planetary nebula KjPn
8. Analytical calculations and numerical simulations are presented
to evaluate the model. The envelope is considered to consist mainly of
environmental gas swept-up by shocks driven by an episodic, collimated, 
bipolar outflow. In this model, which we call the ``free stagnation knot''
mechanism, the swept-up ambient gas located in the stagnation region of 
the bow-shock cools to produce a high density knot. This knot moves along
with the bow-shock. When the central outflow ceases, pressurization of 
the interior of the envelope stops and its expansion slows down. The
stagnation knot, however, has sufficient momentum to propagate 
freely further along the axis, producing a distinct nose at the end of the
lobe. The model is found to successfully reproduce the peculiar shape and
global kinematics of the giant bipolar envelope of KjPn 8.

\end{abstract}

\keywords{hydrodynamics: numerical --- Planetary nebulae: individual KjPn~8---
kinematics and dynamics---jets and outflows}

\section{Introduction}

The planetary nebula KjPn~8 is surrounded by an expanding giant
(14\arcmin $\times$ 4\arcmin) bipolar envelope which so far is unique
in its kind (L\'opez \etal 1995). The physical dimensions of the
envelope as measured along the main axis and along the latest secondary 
outflows have been estimated to be 4.1 $\times$ 1.2~pc, respectively, for
a distance of 1~kpc, corresponding to 0.3~pc~arcmin$^{-1}$
(L\'opez \etal 1995). This size has to be compared with the diameter
of the compact core which is only $\approx$ 4\arcsec\ across. Typical
PNe have sizes of only a fraction of a parsec.

The main bipolar structure presents a tubular shape in the central
region that decreases in radius in a peculiar way at increasing
distance from the central source. The expansion speed of this `tube'
perpendicular to its main axis is $\approx 40$~\kms ~over a large
angular extent and $\approx 160$ \kms ~along it (L\'opez \etal
1997). In addition to its huge dimensions and peculiar morphology, the
bipolar envelope of KjPn~8 presents intriguing ripples across its
structure, some of which mark the locations of strongly decreasing
radius, culminating in the NE lobe in a narrow nose with a bright knot
at the tip. The end of the SW lobe is less well defined but follows a
similar pattern.

The morphology of KjPn~8 has been interpreted as the result of the
action of a bipolar, rotating episodic jet or BRET (L\'opez \etal
1995). There is strong morphological and kinematical evidence that the
ejection direction of the main bipolar outflow has changed with
time. There is at least one additional kinematic subsystem to the main
envelope which seems to be related to more recent ejections at a
projected orientation of $\sim 50^\circ$ with respect to the axis of
the main envelope. Observed radial velocities of symmetric emission
line knots in these regions are as high as $\pm 220$~\kms. The
structure of these secondary bipolar jets suggests
a large opening angle of the outflow, of up to $70^\circ$
(L\'opez \etal 1995; L\'opez \etal 1997). Meaburn (1997) has recently
derived proper motions of individual knots at the heads of these
secondary outflows which combined with a simple bow-shock model yield
a distance of $1600\pm230$~pc for KjPn~8. With this new distance the
angular dimensions transform into 6.5 $\times$ 1.8 pc.

V\'azquez, Kingsburgh \& L\'opez (1998) have obtained low resolution
spectra of the core of KjPn~8 finding it to be a low excitation type I
PN. The nebula seems to have been formed from a massive progenitor
within a metal-rich environment, in agreement with its location right in 
the galactic plane. In addition, spectra of some knots and faint regions
obtained in the nebular envelope indicate very low electron densities, 
ranging from 100 to 300\,cm$^{-3}$. Recently, Huggins \etal (1997) and
Forveille \etal (1998) have detected the presence of a remarkable
expanding molecular CO torus in the core of KjPn~8 whose plane 
is perpendicular to the secondary, high-velocity bipolar outflows.

In this paper we explore a mechanism which we call the ``free
stagnation knot'' to explain the peculiar characteristics of the giant
bipolar envelope of KjPn~8.  In this model the shocked ambient gas in
the stagnation region of the bow-shock of a supersonic jet has
sufficient time to cool in the vicinity of the symmetry axis to form a
dense massive knot, moving at the advance speed of the
bow-shock. After the outflow shuts off, the pressure in the nebula
falls and the expansion speed of the existing envelope drops. The
newly formed dense knot, however, has sufficient momentum to continue
its motion opening a narrow channel along the axis.  Analytical
calculations and time-dependent hydrodynamical simulations are
presented and confronted with the existing observations.

\section{The collimated outflow}

An H$\alpha$ mosaic of KjPn8 (L\'opez et al. 1995) is presented in the
top panel of Figure 1, a blow-up of the north east extreme of the
nebula is shown in the bottom panel of this figure.

The extended envelope of KjPn~8 requires that the density of the jet
be much smaller than that of the ambient medium, a behaviour similar
to that found in extragalactic jets  (Norman, Smarr and
Winkler, 1985). The velocities involved here are, however, only of the
order of a thousand kilometers per second as compared to
velocities close to the speed of light in extragalactic jets. Radiative cooling
will therefore be important, mainly for the shocked ambient medium.
Some aspects of jets with cooling in internal shocks have been
simulated numerically by e.g. Blondin, Fryxell and K\"onigl (1990).
The creation of elongated envelopes from stellar winds propagating
into an inhomogeneous ambient medium has been investigated
semi-analytically by Icke (1988) for the case of the radio nebula W50
around the X-ray binary SS433.  The dynamics of radiative bow-shocks
of continuous adiabatic jets has been investigated by Steffen \etal
(1997, and references therein). These investigations show that an
adiabatic jet results in an overall elongated ellipsoidal or
cylindrical shape of the envelope. Due to quasi-periodic
vortex-shedding in the bow-shock region, it is common to find
large-scale ripples in the envelope superimposed on smaller scale ones
which are caused by instabilities. However, the overall shape always
has a convex curvature.  From the existing studies it is clear that
the cone-like structure of the envelope in KjPn~8 cannot be produced
with a continuous low-density jet without cooling of the jet itself or
a drastic structural anomaly in the bow-shock region.

In order to obtain a baseline for the parameter space for our
numerical model of the envelope of KjPn~8 some basic properties of the
outflow are estimated assuming a continuous non-radiative jet.  The
optically emitting envelope of swept-up ISM gas around the cocoon of a
supersonic jet propagating into a uniform medium is now considered.

The expansion speed transverse to the axis of the main envelope is
$v\u{r} \approx 40$~\kms. The envelope is somewhat asymmetric near the
base (see Figure 1). Assuming axial symmetry, a radius $r\u{0}
\approx 0.4$~pc and a length of $l \approx 2$~pc for the distance from
NE tip to the core is adopted. The radius for the minor axis of the
envelope of 0.4 pc is found by considering the distance from the
core to the filament that traces the envelope located directly south
of the core , thus avoiding the deformation produced by the
high-velocity jets on the envelope, where the radius would amount to
0.6 pc (L\'opez \etal 1995).  Considering that the transverse
expansion speed has been constant over the life-time of the nebula,
its age can be estimated to be $t \sim10^4$~yr. It is, however, likely
to be smaller by a factor of two or so, since the expansion speed must
have been higher in the initial stages of the expansion.

Using as age $t = 10^4$~yr and a propagation distance $l = 2$~pc we
find that the mean advance speed of the bow-shock of the jet is $\sim
200$~\kms. Here we have also assumed that the ejection along the main
axis has stopped only recently on this time-scale (i.e. we ignored the
apparent recent changes in direction of the ejection). This bow-shock
speed is low enough to be in the radiative shock regime, which is
consistent with the observation that the NE lobe of the nebula can be
observed up to the tip of the bow-shock. The SW lobe, however, does
not show a bright tip at its end, possibly indicating that the shocked
ISM collected by the bow-shock has not fully cooled yet.  This could
imply either that the ratio between the density of the jet and ambient
medium has been higher here or that the jet itself had a higher speed
than in the NE half of the nebula. If the age is smaller than
$10^4$~yr, as suggested above, the average advance speed would of
course be higher too.  As a working value for the advance speed of the
outflow we shall therefore adopt $v\u{jh} \approx 400$~\kms.

The observations suggest that there is variability in the direction of
the outflow. Furthermore, the zonal structure of the nebula along the
symmetry axis suggests a possible intermittence of the ejection. Note,
however, that a similar structure can be produced by vortex-shedding
in the bow-shock region or instabilities in the thin shell
(e.g. Steffen \etal 1997).  

The detection of the thin envelope near the end of the nebula requires 
that the mean advance speed of the jet be of the order of a few
hundred kilometres per second. It also follows that the true advance speed
of the active outflow has to be relatively moderate, about 1000~\kms, 
otherwise the cooling time would be too large to make it observable in
\ha\ within the lifetime of the nebula. 
For instance, at an ambient density of 100~\pccm ~and
a bow-shock velocity of 1000~\kms ~the cooling time in the stagnation
region would be around 40000~yr and for lower densities proportionally
higher.  Thus, the advance speed has to be considerably less than
1000~\kms. If the mean advance speed is of the order of 200~\kms\ the
outflow cannot have been switched off for a time significantly longer
than the time it was on during the duty cycle. Otherwise the true
advance speed would have to be too large to provide the observed mean
speed with a cooling time short enough for the formation of the
cool envelope. The cooling time increases
with a high power ($>$3) of the shock velocity, whereas the mean advance
speed is roughly proportional to the ratio between the time that the
jet is switched on and the full duty-cycle. For the overall dynamics
and energetics the possible intermittence will have no large effect on
our estimates, therefore we assume a continuous outflow for the
global estimates of the nebula. Simulations show, however, that
intermittence can have some influence on the overall shape of the
envelope. From the segmented structure of the envelope on a scale of
roughly 0.5~pc for each segment and an estimated bow-shock advance
speed of $\approx$ 400~\kms, the duty cycle is around $4\ee{10}$~seconds.

For our estimates of the physical properties of the envelope and the
jet we assume a conical shape of the nebula and use the dimensions of
the NE lobe. First we estimate the density of the undisturbed medium
into which the nebula expands. We use the measured expansion velocity
of  $v\u{r}=40$\kms, typical mass-loss rates $\dot{M\u{j}}$
between $10^{-8}$ and $10^{-6}~\msol \r{yr}^{-1}$ (Hutsem\'ekers \&
Surdej 1989) and wind/jet velocity in the range $1000-3000$~\kms.

If the kinetic energy of the jet is largely used to accelerate the
mass $M$ of the ambient medium into the thin envelope , then we can
assume conservation of kinetic energy, i.e.
\begin{equation}
\frac{1}{2} \dot{M\u{j}} v\u{j}^2 t = \frac{1}{2} M v\u{r}^2.
\end{equation}
We assume that the envelope is expanding at a velocity of
$v\u{r}=40$\kms\ over the life-time $t$ of the nebula. This yields the
mass of the swept-up ambient medium in the form
\begin{eqnarray}
M    &=& \dot{M\u{j}} t v\u{j}^2  v\u{r}^{-2} \\
     &=& 0.6\msol \frac{\dot{M\u{j}}}{10^{-7} \msol ~ \r{yr}^{-1}} 
         \frac{t}{10^4 ~ \r{yr}} \nonumber \\
      && \cdot \left(\frac{v\u{j}}{1000 ~ \r{km~s^{-1}}}\right)^2 
         \left(\frac{v\u{r}}{40~\r{km~s^{-1}}}\right)^{-2}. 
\end{eqnarray}
For simplicity we assume a conical outline of the nebula and a uniform
ambient medium of number density $n$, which is then calculated from
\begin{eqnarray}
\label{mass.eq}  
n &=& \frac{3}{\pi} \frac{M}{m_\r{p}} l^{-1} r\u{0}^{-2} \\
  &=& 80~\r{cm}^{-3} \left(\frac{M}{0.6\msol}\right)
         \left(\frac{l}{2~\r{pc}}\right)^{-1}
         \left(\frac{r\u{0}}{0.4~\r{pc}}\right)^{-2},
\nonumber
\end{eqnarray}
where $m_\r{p}$ is the mass of the hydrogen atom (we assume a pure
hydrogen nebula).  Since $M$ is the mass of only one side of the
envelope, complete envelope will have a mass of roughly $2M$. Via the
size scale the density $n$ is inversely proportional to the cube of
the distance of KjPn~8. Using the distance of 1600~pc as determined by
Meaburn (1997), the density of the ambient would be 330\pccm. Since
the cooling time are inversely proportional to the density, an
accurate determination of the distance to KjPn~8 is important for our
model.

The number density $n\u{j}$ of the collimated jet can then be
estimated from the ratio $\zeta$ between the jet velocity $v\u{j}$ and
the bow-shock speed $v\u{jh}$. From ram-pressure balance at the
working surface of a supersonic jet, we have
\begin{equation}
\label{nj.eq}
n\u{j} = n (\zeta - 1)^{-2} .
\end{equation}
For a typical range of wind/jet velocities given above and the roughly
estimated advance speed $v\u{jh}=400$\kms, $\zeta$ is in the range 2.5-7.5.
This translates into a jet density between about 2 and 35~\pccm ~for
$n=80$\pccm.

The radius of the jet after collimation is then
\begin{eqnarray}
\label{jetrad.eq}
 r\u{j} &=& \left(\frac{1}{\pi \r{m_p}} \dot{M} 
            n\u{j}^{-1} v\u{j}^{-1}\right)^{1/2} \\
        &=& 3.5\ee{16}~\r{cm}~ 
            \left(\frac{\dot{M\uj}}{10^{-7}~\msol\r{yr}^{-1}} \right)^{1/2} 
            \nonumber \\
        &&  \left(\frac{n\u{j}}{10~\r{cm}^{-3}}\right)^{-1/2} 
            \left(\frac{v\u{j}}{1000~\r{km~s^{-1}}}\right)^{-1/2} .
            \nonumber 
\end{eqnarray}

Using the range of mass loss, jet velocity and densities estimated
above, the radius of the jet after collimation is within approximately
one order of magnitude of $3.5\ee{16}~\r{cm}$.  This radius is similar
to that of the CO molecular ring found by Huggins \etal (1997) and
Forveille \etal (1998). The axis of the ring is aligned with the most
recent high velocity jets. The possible relation of this toroid with a
similar structure related to the origin of the bipolar large envelope
is at present uncertain.  However, it is interesting to note that this
peculiar situation is not unique among PNe. Another example is found
in the multipolar PN NGC 2440 (L\'opez et al. 1998) where HST images
also reveal a toroidal structure at its core. The plane of this ring
is not orthogonal to the axis of the main bipolar structure either.
This is a problem in PN evolution that has not been addressed yet in
any detail and clearly deserves further investigation.  For this
analysis we shall simply assume, without further consequences for the
models, that a high density equatorial ring has been related to the
collimation of the outflow, in a process similar to the one described
by Mellema \& Frank (1997).
 
For the formation of an extended envelope from a low-density jet a
Mach number $\r{M}>5$ is required (Norman, Smarr and Winkler
1985). A Mach number $\r{M}=10$ yields a temperature of the
jet given by
\begin{eqnarray}
T\u{j} &=& \frac{\bar{m}}{\gamma k}\frac{v\u{j}^2}{\r{M}^2} \\
       &=& 3.6\times10^{5}\r{K} \left(\frac{\r{M}}{10}\right)^{-2}
         \left(\frac{v\u{j}}{1000~\r{km~s^{-1}}}\right)^{2}. \nonumber
\end{eqnarray}
Here $\gamma=5/3$ and $\bar{m}$ is the mean molecular weight and $k$
is the Boltzmann constant (we assume $\bar{m}=0.5\mpr$, $\mpr$ is the
proton mass). For jet velocities of up to 3000\kms\ and a Mach number
fixed at $\r{M}=10$, this yields a jet temperature of up to a few
times $10^6$~K.

A jet of this temperature and velocity is capable of producing a
significant amount of thermal X-ray radiation. The highest intensity
can be expected to emerge from the hot shocked jet gas in the
stagnation region of the active jet. Diffuse emission should be
found in the large volume of the still hot cocoon region. The
emissivity $\epsilon\u{x}$ from bremsstrahlung can be calculated from
(Cox and Tucker, 1969)
\begin{equation}
\epsilon\u{x} = 2.3\ee{-27} \left(\frac{n}{1{\rm cm}^{-3}}\right)^2 
                         \left(\frac{T}{1\rm{K}}\right)^{1/2},
\label{xray.eq}
\end{equation}
where $T$ and $n$ are the temperature and number density of the
emitting gas, respectively.  At a pre-shock velocity of 3000\kms\ the
post-shock temperature will be $10^8$K. For an order of magnitude
estimate of the expected X-ray emission in the stagnation region let's
assume $n=4n\uj=40$\pccm\ and an emitting volume $V=4\pi r\uj^3/3 =
4\ee{51}\r{cm}^{-3}$ ($r\uj=1\ee{17}\r{cm}$). This yields a total
luminosity $L\u{x}=1.5\ee{32}
\r{erg ~sec}^{-1}$. At a distance of 1~kpc this corresponds to an 
integrated flux of $1.3\ee{-12} \r{erg ~sec}^{-1}\r{cm}^{-2}$ which would 
be within reach of the AXAF space  telescope. Obviously, X-ray observations 
of KjPn~8 will be an important test of some aspects of our jet model.

\subsection{Summary of estimated parameters}

The estimated parameters for the envelope of KjPn~8 and the outflow,
which has produced it, are:
\begin{eqnarray}
t      &\approx& 10^{4}~\r{yr}                 \nonumber\\
n      &\approx& 80~\r{cm}^{-3}                \nonumber\\
M      &\approx& 0.6~\msol                     \nonumber\\
n\u{j} &\approx& 2~-~35~\r{cm}^{-3}            \nonumber\\
r\u{j} &\approx& 3.5\ee{15-17}~{\rm cm}        \nonumber\\
v\u{j} &\approx& 1000-3000~\r{km~s}^{-1}       \nonumber\\
v\u{jh}&\approx& 400~\r{km~s}^{-1}             \nonumber\\
T\u{j} &\sim   & 10^6 K                        \nonumber
\end{eqnarray}
The parameters are not all independent from each other. For instance,
the advance speed $v\u{jh}$ of the bow-shock has been estimated from
the expansion speed of the envelope and its size (the latter in turn
depends on the distance).  Keeping the advance speed fixed, while
changing the ambient density requires a corresponding change of the
jet density or velocity. These changes of the jet parameters have
strong implications for the cooling of the jet and thereby on the
local radii of the jet and the envelope.

\section{The free stagnation knot model}

We now consider the required time-scales for the formation of the
stagnation knot and estimate its basic properties like density and
size. A schematic illustration of this mechanism is presented in
Figure 2. The top diagram shows the out-flowing jet with the extended
expanding envelope of shocked ambient gas. In the stagnation region a
dense knot has formed from cooling shocked interstellar gas and
propagates along with the bow-shock. After the outflow ceases (bottom)
the pressure inside the envelope decreases and becomes more
uniform. During this stage the stagnation knot continues to move along
the axis and forms a narrow nose to the wide envelope.

For the formation of a dense knot in the stagnation region of the
bow-shock, the cooling time of the shocked ambient medium in this
region has to be smaller than the age of the outflow. 

Taking the
average advance speed $v\u{jh}$ of the jet head as the shock speed, 
the post-shock cooling time in the stagnation region is
\begin{equation}
\label{cooltime.eq}
t\u{cl} =  2450~\r{yr} \left(\frac{n}{80~\r{cm}^{-3}}\right)^{-1}
          \left(\frac{v\u{jh}}{400~\r{km s}^{-1}}\right)^{3.26}.
\end{equation}
Here we have used a cooling function of the form (e.g. Blondin,
Fryxell and K\"onigl)
\begin{equation}
\label{fcool.eq}
\Lambda (T) = \Lambda_0 T^\alpha
\end{equation}
with $\Lambda_0 = 1.05 \cdot 10^{-18} {\rm erg\,s}^{-1}$\pccm\ and
$\alpha=-0.63$ at $T > 1.5 \cdot 10^5$K. These values closely describe
the cooling in the corresponding temperature regime in the
numerical code used for our simulations. The temperature $T$
immediately behind the shock is given by
\begin{equation}
\label{shocktemp.eq}
T = \frac{3}{16} \frac{\bar{m}}{k} v\u{jh}^2.
\end{equation}

The outflow must have lasted and kept its direction for a period
longer than this value. Otherwise the gas in the stagnation region
might not cool before flowing off into the cocoon. The cooling time
$t\u{cl}$ thereby represents a lower limit to the duty cycle of any
intermittence of the outflow. The value is however rather uncertain,
due to the strong dependence on the advance speed of the bow-shock
$v\u{jh}$, which is not well known for KjPn~8. A higher ambient
density $n$ would make the process more efficient, reducing the
cooling time proportionally. Even within the uncertainty of a factor
of two for the bow-shock speed, the stagnation knot should be formed
quite early within the estimated age of the nebula ($t<10^4$~yr).

As long as the knot does not brake up and is highly supersonic
(as is the case here) it is ``free'' to continue its motion through the
ambient medium until it has swept up roughly as much mass as its
own. The distance $d$ which it will reach by then is
\begin{equation}
\label{stopknot.eq}
d = \frac{4 n\u{k0}}{3n} r\u{k0},
\end{equation}
where $n\u{k0}$ and $r\u{k0}$ are the number density and the spherical
radius of the knot, respectively. The density is estimated by assuming
that the knot is in pressure equilibrium with the surrounding medium
in the stagnation region, which has roughly the same thermal pressure
as the ram pressure of the jet. For the final temperature of the knot
it appears reasonable to assume roughly $T=10^4$~K. The number density
of the knot is then
\begin{eqnarray}
\label{knotdensity.eq}
n\u{k0} &=& n\uj \frac{\bar{m} v\uj^2}{kT} \\
        &=& 6\ee{4}~\r{cm}^{-3} \frac{n\uj}{10\r{cm}^{-3}} 
          \left(\frac{T}{10^4~\r{K}}\right)^{-1}
          \left(\frac{v\uj}{1000~\r{km~s}^{-1}}\right)^2 \nonumber
\end{eqnarray}
where $\bar{m}$ is the mean atomic mass (assumed to be half a hydrogen
atom to account for ionization) and $k$ is the Boltzmann constant.
Using the values estimated in Section 2 the density in the knot is
$n\u{k0} = 4.1\ee{5}$~\pccm.  We find an upper limit for its radius by
assuming that only ambient gas can cool which is located within a jet
radius $r\uj$ of the axis. Any gas further away from the axis will
flow into the cocoon. For a knot that has been compressed to the density
$n\u{k0}$ it is then found that the spherical radius $r\u{k0}$ is given by
\begin{equation}
\label{knotradius.eq}
\pi r\uj^2 v\u{jh} t n  > \frac{4\pi}{3} r\u{k0}^3 n\u{k0},
\end{equation}
where $v\u{jh}$ is the advance speed of the bow-shock and $t$ is the
age of the source at the time when the cooling in the stagnation
region becomes significant.  Using Equation \ref{nj.eq} in
the approximation of a light jet ($\zeta\gg 1$) and combining it with
Equations \ref{knotdensity.eq} and \ref{knotradius.eq}, the upper
limit for the radius of the knot is
\begin{eqnarray}
\label{knotradlim.eq}
r\u{k0} &<& \left(\frac{3}{4} r\uj^2 \frac{t}{v\u{jh}} 
          \frac{kT}{m}\right)^\frac{1}{3} \\
        &=& 4.6\ee{16}~\r{cm} \cdot \nonumber \\
         &&  \left[\left(\frac{r\uj}{10^{17}\r{cm}}\right)^2
            \frac{t}{10^4~\r{yr}}
            \frac{T}{10^4~\r{K}}
            \left(\frac{v\u{jh}}{400~\r{km~s}^{-1}}\right)^{-1}
            \right]^\frac{1}{3}. \nonumber
\end{eqnarray}
Applying the estimates for  $n\u{k0}$ and $r\u{k0}$ to Equation
\ref{stopknot.eq} we find $d\us=4.6\ee{19}$~cm~$=15$~pc, which is
interpreted as an upper limit to the distance the stagnation knot can
travel. A lower limit for this distance can be estimated by
considering the expansion of the knot.  As long as the outflow is
active, the stagnation knot will be confined by the pressure in the
bow-shock region and advance at the bow-shock speed. Once the outflow
ceases the pressure will fall, allowing the stagnation knot to
increase its radius roughly with its internal sound speed
$c\u{s}$. The sound crossing time corresponding to
$r\u{k0}=4.6\ee{16}~\r{cm}$ is $t\u{s}=3.8\ee{10}$~s.  If the knot
moves at $v\u{jh}=400$\kms, then it will travel a distance $d\us =
1.2\ee{18}$~cm~$\approx 0.4$~pc in this time.  Assuming that the knot
can survive a few times the sound crossing time before it fully
disintegrates and stops, then the knot can travel for a
distance of the order of 1~parsec (for the typical parameters used in
Equation \ref{knotradlim.eq}). These values are consistent with the
observed distance of 2~parsec between the central source and the tip
of the envelope of KjPn~8.

\section{Numerical simulations}

The numerical simulations have been carried out with the adaptive grid
hydrodynamic code described by Raga (1994) in axisymmetric and
slab-symmetric mode.  This code solves the equations of mass, momentum
and energy conservation using a flux-vector-splitting scheme (van Leer
1982).  The computations were carried out on a 5-level, binary
adaptive grid. The maximum grid size was 1025$\times$513 and
513$\times$513 cells in the case of axial and plane symmetries,
respectively. The non-equilibrium cooling as described in Biro, Raga
and Cant\'o (1995) has been used.  For low temperatures, energy loss
from the collisional excitation of [O I] and [O II] lines and
radiative recombination of H have been taken into account.  At
temperatures higher than $5\ee{4}$K, a parameterised coronal
equilibrium cooling rate is used. The collisional ionization of H and
excitation of Lyman-alpha are also included in the cooling.
The boundary conditions were reflective on the axis and on the left
side of the computational domain (except for the inflow condition
where the jet is injected). For the top, bottom (in plane symmetry)
and right boundaries outflow conditions were applied.  The jet is
injected with uniform velocity and density over its radius.  The
\ha\ emissivity has been calculated using radiative
recombination (Case B) and collisional excitation following Aller
(1984). 

The slab-symmetric simulation presented in the next section is used to
illustrate the qualitative details of the formation of the stagnation
knot while excluding possible singular numerical effects on the axis
of the axisymmetric computation. Parameters of the slab-symmetric
simulation have been chosen for clarity of illustration of the
mechanism of the formation of the stagnation knot, rather than for
comparison with KjPn8.  For a more detailed comparison with the
observations of KjPn8, we use an axisymmetric model, since it provides
a more realistic calculation of the off-axis kinematics and the
emission of the envelope, as well as for the compression and
kinematics of the free stagnation knot.

\subsection{Results and discussion}

A series of simulations was performed in which we
investigated the effects of varying the densities of the jet and the
ISM ($n\uj=0.1-20$\pccm; $n\u{ISM}=1-150$\pccm), the jet velocity
($v\uj=500-4000$~\kms), its initial radius
($r\uj=0.5-3\ee{17}\r{cm}$), the jet half opening-angle ($0-25^\circ$)
and the grid resolution (up to 1025$\times$513 grid points at the
lowest level of the adaptive grid). As a final step we studied the
effect of jet pulsation on the set of parameters that appeared to fit
best the characteristics of KjPn~8. To ensure sufficient resolution
over the jet radius, radii smaller than $0.5\ee{17}\r{cm}$ were not
investigated.

The tests on the grid showed that at the highest
resolution only the quantitative details of the simulations were still
somewhat dependent on the resolution, especially those relevant to the
cooling of the jet and the stagnation knot.  Also, the detailed
structure of the instabilities in the cocoon envelope slightly changed
with resolution at early times, while it was still confined by the
over-pressured cocoon. Increasing the resolution above the
1025$\times$513 used in the simulations shown in this paper would
have led to prohibitively high computing times.

The intermittence of the jet on a time-scale of around 1000~yr adds to
the quasi-conical shape of the nebula as opposed to a more cylindrical
shape obtained for a continuous ejection of the same jet. However, it
is not clear from the simulations what exactly determines how the
intermittence changes the shape.  Relevant quantities could be the
on/off ratio or the ratio between the times of the duty-cycle and the
quasi-periodic vortex shedding near the head of the jet. The
intermittence also emphasizes the division of the envelope into a few
sections of different radii separated by rings of higher
emissivity. In Section 2 we used this to estimate the duty-cycle of
the episodic jet.

In the following, we describe representative simulations with very
different jet parameters which all show the formation of the
stagnation knot. These include one slab-symmetric run for a detailed
demonstration of the formation process. The other two simulations are
performed using cylindrical symmetry, one with a fully collimated jet
and the other with a large opening angle. The parameters of
these runs are listed in Table 1.
\begin{table}
\label{runs.tab}
\begin{tabular}{llrrr}
                  &            & COLLIMATED    &   UNCOLLIMATED  
                  & SLAB       \\
1/2 opening angle &  $\theta$  & 0             &   25           
                  & 5          \\
jet radius        &  $r\uj$    & 0.1           &   0.15         
                  & 0.075      \\
jet velocity      &  $v\uj$    & 3000          &   4000          
                  & 1100       \\
jet density       &  $n\uj$    & 3             &   1.25          
                  & 30         \\
jet temperature   &  $T\uj$    & $2\ee{6}$     &   $2\ee{6}$    
                  & $2\ee{5}$  \\
mass loss rate    &  $\dot{M}\uj$ & $7.4\ee{-7}$ &   $7.4\ee{-7}$ 
                  & $7.7\ee{-7}$\\
ISM density       &  $n   $    & 100           &   25           
                  & 80         \\
ISM temperature   &  $T   $    & $10^4$        &   $10^4$         
                  & $10^4$     \\
domain of simul.  &            & $12\times1.5$ &   $10 \times 5$  
                  & $3\times1.5$\\
on/off times      &            & 0.3/0.1       &   ----          
                  & ----       \\
cut-off time      &            & 1.2           &   3.2       
                  & ----     

\end{tabular}
\caption{The parameters of the representative runs showing the 
stagnation knot with a fully collimated jet, an outflow with a 
large opening-angle of 25$^\circ$ and a slab-symmetric run with 
a small opening-angle. Lengths are given in
units of $10^{18}$\,cm and times are in units of $10^{11}$\,sec.
Number densities are given in units of 1\,\pccm, velocities in
\kms and temperatures in Kelvin. The mass loss rate is given 
in solar masses per year. }
\end{table}

Figure 3 shows several stages of the formation of the stagnation knot
during the slab-symmetric run. The images are grey-scale
representations of the density (Fig. 3, Panels a,b and d) and pressure
(Figure 3, Panel c) with velocity arrows superimposed.  Panels b and c
of Figure 3 are density and pressure representations for the same time
during the simulation. Only the bow-shock section from the larger full
domain of the calculation is shown. Note that the velocity vectors are
shown in the reference frame moving along with the bow-shock at speed
$v\u{jh}$ as calculated from ram-pressure balance (equivalent to
Eq. \ref{nj.eq}), i.e.
\begin{equation}
\label{vjh.eq}
v\u{jh} = \frac{v\uj}{1+\sqrt{\frac{n}{n\uj}}}.
\end{equation}
This changes the velocity vectors in the stagnation region to zero
length.  The validity of this condition can be appreciated in 
Figure 3, Panel a, where the velocity vectors in the stagnation region effectively have zero-length. Transforming the velocities to this
system makes it easier to note the changes imposed by the cooling of
the stagnation region.  For clarity, vectors corresponding to
velocities higher than 300\kms\ in this frame have been omitted. The
longest velocity vectors in the immediate post-shock region of the
bow-shock correspond to 0.25$v\u{jh}$ which is approximately
100\kms. Initially, the post-shock region of the ISM is roughly
uniform in density (t=700~yr, Fig. 3, Panel a). All the velocity
vectors point along or diverge from the axis of the jet.  At t=1200~yr
the regions of highest density begin to cool noticeably and after 60
more years a significant increase in density and a decrease in
pressure is seen (Panels b and c, respectively). 

The reduction in pressure in the stagnation region causes a dramatic 
change in the structure of the bow-shock, shaping it concave instead 
of convex near the axis (Fig. 3, Panel d). 
In the concave region the swept-up ISM is now refracted towards the 
stagnation region thereby feeding the newly formed cold knot with fresh material. As can be seen from the short vectors associated with the cold 
knot, its velocity is very well reproduced by Equation \ref{vjh.eq}. 
Given the reduced pressure, material is being accelerated towards the cold
dense plug.
This is best seen in Figure 3, Panel d, where both shocked gas from the jet
and the ISM converge towards the stagnation knot from both sides. This
is indicated by the presence of velocities of the order 100~\kms\ in the
frame of the stagnation region (as defined by Equation
\ref{vjh.eq}). 

The process observed here is similar to the formation
of a nose cone in the case of a jet which is denser than the ambient
medium.  In that case the condensing material is from the jet which
is refracted towards the axis in a concave jet-shock (Raga, Cant\'o \&
Cabrit 1998). Note that shortly after cooling the stagnation knot
already shows signs of instability and fragmentation, which is
important for its future development after the jet ceases and the
condensation is set free. As discussed before, the compactness and
therefore the degree of fragmentation strongly determines the distance
the free stagnation knot can travel.

The formation of a stagnation knot in the axisymmetric run can be
observed in Figure 4. The condensation moves outwards at the bow-shock
speed, remaining at the boundary of the large-scale nebula. In this
simulation the dense knot started to grow after approximately 1250
yr. Despite of the very different jet parameters, the cooling time is
similar to the one found in the slab-symmetric simulation, since the
advance speed of the bow-shock is approximately the same in both
cases.  The stagnation knot remains confined until the jet is switched
off (at $t=3470$~yr, after a period of intermittence with a duty-cycle
of $t=1260$~yr). After this time the knot starts to expand as the
pressure in the cocoon drops (Figure 5). The high momentum keeps the
knot propagating freely into the ambient medium at the original speed
of the bow-shock for some time. Consequently, it starts to speed ahead
of the original wide bow-shock, ploughing a narrow channel or `nose'
into the ISM. At the same time it increases its radius and flattens
into a pancake-shaped slab. Instabilities cause it to form ripples
which develop into individual smaller knots, resulting in a break-up
of the original stagnation knot.  The advance speed of the knot
clearly decreases and an extrapolation shows that it would stop after
traveling a distance of at most 1.5~pc from the point at which the jet
was switched off ($\approx$ 2.5~pc from the source). This value is in
good agreement with the estimates in Section~4 and with the position
of the observed end of the nebula in KjPn~8.

Figure 6 shows the time evolution of the \ha\ emission from the
simulated nebula from $t=2840-9150$~yr as seen from a viewing angle
almost perpendicular to the axis. It shows how the stagnation knot is
compact at the beginning and then becomes diffuse as the pressure is
not maintained by the jet anymore. In the last image the shape of the
whole nebula is very similar to the envelope around KjPn~8 (Figure
1). The stagnation knot appears as a flat tip of the narrow nose
very much like the observed one.  Especially the north eastern lobe of
the nebula is very well reproduced by the simulation, with the bright
tip at the end associated with the stagnation knot. The surface
brightness of the inner bulge of the nebula is higher than in the
sections were the axial radius drops markedly. This is also found in
the simulation, together with a good reproduction of the ripples
caused by instabilities and intermittence of the jet.

After the jet injection ceases the pressure in the cocoon slowly
decreases and the expansion of the envelope near the base slows down
rather quickly below 50\kms. In Figure 7 the position-velocity (pv)
diagrams of the H$\alpha$ emissivity are shown (top and right
panels). They correspond to the final stage of the time-series in
Figure 6.  In these diagrams position runs along the white dotted
line, which also marks the location of zero expansion velocity.
Positions in the pv-diagrams project directly onto the H$\alpha$
image, also shown (lower left).  The diagrams represent thin slices
through the cylindrical simulation.  The ``slit'' in the top diagram runs
along the axis, while the other runs perpendicular to the axis (but
looking along it).  Only one of the two symmetric sections in the cut
through the cylindrical envelope have been shown (exactly as
represented by the H$\alpha$ image). 

The expansion velocities at a time around 6000~yrs - corresponding to
the fifth panel from the top in Figure 6 - are consistent with those
measured for KjPn~8. After that, they fall below the observed value of
40\kms\ to approximately 25\kms.  In the top pv-diagram of Figure 7
the velocity pattern drastically changes at the position where the
inner wide envelope ends and the nose caused by the stagnation knot
starts. This kinematic signature could be used as a further test
performing spectral observations aiming at mapping the velocity field
in the the corresponding area of KjPn~8. The currently available
observations do not cover this area suitably.

The formation of the stagnation knot may be favoured by the
axi-symmetric nature of the simulations. In this symmetry no change in
direction of the outflow is allowed by possible jet-instabilities or
deflections due to pressure variations in the cocoon. Changes of the
flow direction would allow the gas in the stagnation region to flow
off into the cocoon. The formation of the stagnation knot is therefore
most likely to occur before the first internal reflection shock of the
jet forms if the cooling time in the stagnation region is sufficiently
short. In fact, even if the bow-shock speed was small enough, in none
of the simulations performed a stagnation formed if a reflection shock
was present before the stagnation region had time to cool.  Instead,
the whole bow-shock region would cool, producing a thin shell rather
than a substantial axial knot.  We attribute this to the shape of the
bow-shock, which does not become concave, even after the cooling
starts. The reason for this may be that, after the reflection shock,
the radial momentum distribution in the jet peaks on  the axis. This
prevents the formation of a concave working surface.

If the outflow has a substantial opening angle, it is less sensitive
to instabilities which temporarily could change its flow direction. In
Figure 8 we present a simulation with an outflow which is poorly
collimated and does not recollimate.  Here the working surface and the
stagnation region are larger compared to more collimated jets,
resulting in a longer time available for cooling. Bow-shock velocities
of flows with high opening angles decrease with increasing size of the
shocked region, resulting in smaller cooling times for the shocked
ISM. This further favours the formation of a stagnation
knot. However, the knot has to form in the early stages of expansion
before the bow-shock becomes fully radiative, otherwise only a thin
convex shell is formed.  Thus, a large finite opening-angle has
interesting effects as it can improve the conditions for the formation
of a stagnation knot if the jet does not recollimate.  This occurs
from a half-angle of approximately $20^\circ$ on (Falle 1991, Peter \&
Eichler 1995). In such a case the shape of the envelope adopts a
rather ``boxy''structure when viewed from an angle perpendicular to
the cylindrical axis (see Figure 8). This structure is similar to
those found in some proto-planetary nebulae on a smaller scale, which
so far have lacked a natural explanation (Bryce \etal 1997). Although
in the present paper this is not explored further, a similar model
might apply to these ``boxy'' PNe.

The extraordinary nature of the planetary nebula KjPn~8 appears to be
linked to its environment. The quantitative conclusions from our model
depend on the density of the ISM right before the formation
of the nebula. Density variation over the size-scale of the envelope may 
be responsible for some of the asymmetries found in the giant envelope. 
For instance, the absence of a bright tip at the end of the SE arm could be
related to a strong decrease in density. Sensitive searches of PNe embedded 
in diffuse clouds in the galactic plane might turn out more similar cases of
wind/environment interactions.

Although the main focus of this work has been on the dynamical
modeling of the giant envelope of KjPn~8, it is nevertheless of
interest to add a note on the possible nature of the progenitor of
KjPn~8.  The nebula is outstanding among PNe and as such an uncommon
origin may be expected. One clue may lie in the ionic abundances of
the nebular core derived by V\'azquez, Kingsburgh \& L\'opez
(1998). These authors find that KjPn~8 is an extreme type I PN with
remarkably high ratios of He/H and Ne/O and similar to those found in
He~2-111, another giant bipolar PN with high expansion velocities
(Meaburn \& Walsh 1989).  PNe of type I are produced by massive
progenitors and are generally bipolars (Peimbert \& Torres-Peimbert
1983; Corradi \& Schwarz 1995). Furthermore, Yungelson, Tutukov \&
Livio (1993) have discussed the type of binary systems that may lead
to He rich PN envelopes which also involve massive progenitors that
produce CO or ONe white dwarf nuclei. Coalescence of the binary
nucleus during the AGB stage and more than one common envelope event
and envelope ejection are possible paths in their analysis.

PNe are formed from stars with ZAMS masses $0.8 \lesssim M/M_{\odot}$
$\lesssim 10$, however, most PNe show remnant cores of $\sim 0.6$
M$_{\odot}$, corresponding to progenitors of around 2 - 3
M$_{\odot}$. PNe with high-mass progenitors form spectacular and
complex envelopes such as NGC 2440 and NGC 6302; these are bipolar
type I PNe with core masses estimated around 0.8 - 0.9 M$_{\odot}$ and
must have originated from progenitors of $\sim$ 6 - 7 M$_{\odot}$
(e.g. Pottasch 1983).
 
Thus, with these considerations, a reasonable possibility, although at
this stage necessarily speculative, is to consider that objects like
KjPn~8 and He~2-111 may have their origin in extremely massive
progenitors, those located in the high-mass tail of the distribution
of objects that produce PNe. This type of objects can be expected to
be rare for in addition to the intrinsic lower number of available
progenitors, their evolution as PNe would be very fast across the H-R
diagram.

One further point that is worth mentioning is the presence of the
high-velocity jets in KjPn~8 oriented at a substantially different
position angle from the main symmetry axis of the main bipolar
envelope.  This secondary and younger jet system could in principle be
interpreted as the result of some sort of rotation of the symmetry
axis, as has been done in the case of Fleming 1 (L\'opez, Meaburn \&
Palmer 1993) and numerically simulated by Cliffe et
al. (1995). However, as discussed in Section 2 of this paper, the
recent discovery of an expanding molecular ring or toroid whose plane
is perpendicular to the high-velocity jets and consequently far
off-axis from the main bipolar envelope (Forveille et al. 1998),
represents now a severe obstacle for that interpretation. The
analogous case of NGC 2440 has also been mentioned. In these cases it
becomes interesting to consider the possibility of a recurrent
envelope ejection after either coalescence of a binary nucleus,
additional common envelope event, thermal flash or fast evolution of a
secondary component. The resultant symmetry axis of the bipolar
outflow and presumably the associated dense toroidal shells may have
a different orientation in this occasion due to the dynamical
perturbations occurring in the core. Within this picture, the cases
where secondary jet systems are related to toroidal structures -  which
in turn are tilted with respect to the original bipolar envelopes - may
possibly be understood. Otherwise the confronting elements are
difficult to reconcile within our present limited understanding of PN
formation and evolution.  

\section{Conclusions}

A hydrodynamical model for the formation and evolution of the
peculiar, giant, bipolar envelope of the planetary nebula KjPn~8 has
been investigated in analytical and numerical form.  In this model a
dense knot is formed from the shocked interstellar medium in the
stagnation region of a supersonic episodic jet if the cooling time is 
smaller than the dynamical time needed for the shocked ISM to flow off 
into the cocoon of the jet. During the quiescent phase the knot conserves
enough momentum and continues propagating freely, producing a distinct
nose and a bright tip at the end of the lobe. From the analytical and 
numerical calculations it is concluded that the free stagnation knot model
successfully reproduces the peculiar shape and kinematics of the giant 
envelope of the extraordinary planetary nebula KjPn~8.

\section{Acknowledgements}
We thank Alejandro Raga for permission to use and modify his
hydrodynamic code CORAL and Jose Luis G\'omez for providing the
software for the volumetric rendering of the axisymmetric emissivity
data of the simulations. WS acknowledges the receipt of a PPARC
research associateship and travel support from UNAM-DGAPA grant
IN11896 for a visit to the IAUNAM-Ensenada during which part of this
research was done. JAL acknowledges financial support from UNAM-DGAPA
projects IN11896 and 101495. The authors acknowledge fruitful
discussions with J. Meaburn, M. Bryce and G. Garc\'{\i}a-Segura. We
thank the referee, Adam Frank, of stimulating suggestions that
improved the presentation of this work.

\clearpage

\figcaption{Top: An \ha\ mosaic of the giant ($14^\prime$$\times 4^\prime$) bipolar envelope 
of the planetary nebula KjPn~8 (L\'opez \etal 1995).
Bottom: Enlargement of the NE region.
\label{fig1}}

\figcaption{Schematic view of the stagnation knot scenario before (top)
and after the outflow stops.
\label{fig2}}

\figcaption{A slab symmetric simulation of the formation of the  
stagnation knot. Only the section of the bow-shock of a larger
area of simulation is shown, which comprises the complete region
affected by the interaction of the jet with the interstellar medium.  
\label{fig3}}

\figcaption{Evolution of the logarthmic density distribution  
of the axi-symmetric run  starting at time $t=1260$~yr and
ending at $t=8830$~yr. The pairs of logarithmic minimum and maximum 
for the individual images are (from top to bottom): (-0.9, 4.6),
(-1.5,4.6), (-1.7, 6.1), (-1.7, 4.1), (-1.9, 4.2), (-1.9, 3.9).
\label{fig4}}

\figcaption{Logarthmic pressure distribution  
corresponding to the density distribution shown in Figure 4.  The
pairs of logarithmic minimum and maximum for the individual images are
(from top to bottom): (-10.1, -7.0), (-11.0, -6.9), (-11.7, -5.7),
(-11.5, -7.5), (-11.0, -7.4), (-11.4, -7.7).
\label{fig5}}

\figcaption{Volumetric rendering of the \ha-emission from the simulation
of the stagnation knot scenario (log-scale). The interval between the
individual frames is 1578~years, beginning at 2840~years after the 
start of the simulation. Note the deceleration of the stagnation
knot, soon after it expands and breaks up.
\label{fig6}}

\figcaption{Position-velocity diagram of the envelope for a slit running
along the axis of the envelope (top) and perpendicular to it, looking
down the axis (right). The dotted lines mark zero expansion velocity.
The emission of the stationary external medium has been suppressed in
the pv-diagrams. The bottom left picture shows the logarithmic
H$\alpha$ emissivity distribution as a positional reference to the
pv-diagrams. In the pv-diagrams position runs along the white dotted
line. The axial length-scale has been stretched by a factor of two as
compared to the axial scale (see scale indicators in image).
\label{fig7}}

\figcaption{The logarithmic density distribution of the nebula and stagnation
knot formed from an ouflow with an initial half-opening angle of $25^\circ$. 
\label{fig8}}

\end{document}